\documentclass[sigconf]{acmart}

\AtBeginDocument{%
  }

\usepackage{xcolor} 
\usepackage[most]{tcolorbox} 
\definecolor{lightblue}{rgb}{0.2, 0.6, 0.9}
\definecolor{lightred}{rgb}{0.9, 0.4, 0.4}
\definecolor{lightyellow}{rgb}{0.8, 0.8, 0.35}
\definecolor{lightpurple}{rgb}{0.6, 0.5, 0.7}
\definecolor{lightorange}{rgb}{1.0, 0.5, 0.0}
\definecolor{lightgrey}{rgb}{0.5, 0.5, 0.5}

\usepackage{geometry}
\usepackage{multirow}
\usepackage{booktabs}
\usepackage{longtable}
\usepackage{makecell}
\usepackage{array}

\renewcommand{\arraystretch}{1.2}

\newcommand{\threerowcell}[1]{\makecell[l]{\parbox[t]{14cm}{\renewcommand{\arraystretch}{1.3}#1}}}

\setcopyright{acmlicensed}
\copyrightyear{2024}
\acmYear{2024}
\acmDOI{XXXXXXX.XXXXXXX}

\acmConference[Conference acronym 'XX]{Make sure to enter the correct
  conference title from your rights confirmation emai}{June 03--05,
  2024}{Woodstock, NY}
\acmISBN{978-1-4503-XXXX-X/18/06}

\begin{document}

\title{ECC Analyzer: Extract Trading Signal from Earnings Conference Calls using Large Language Model for Stock Volatility Prediction}

\author{Yupeng Cao}
\email{ycao33@stevens.edu}
\authornote{Equal Contribution}
\affiliation{%
  \institution{Stevens Institute of Technology}
  \city{Hoboken}
  \state{NJ}
  \country{USA}}

\author{Zhi Chen}
\authornotemark[1]
\email{zchen100@stevens.edu}
\affiliation{%
  \institution{Stevens Institute of Technology}
  \city{Hoboken}
  \state{NJ}
  \country{USA}}

\author{Qingyun Pei}
\authornotemark[1]
\email{qpei1@stevens.edu}
\affiliation{%
  \institution{Stevens Institute of Technology}
  \city{Hoboken}
  \state{NJ}
  \country{USA}}

\author{Nathan Jinseok Lee}
\email{nathanlee@hunterschools.org}
\affiliation{%
  \institution{Hunter College High School}
  \city{New York}
  \country{USA}}

\author{K.P. Subbalakshmi}
\email{ksubbala@stevens.edu}
\affiliation{%
  \institution{Stevens Institute of Technology}
  \city{Hoboken}
  \state{NJ}
  \country{USA}}

\author{Papa Momar Ndiaye}
\email{pndiaye@stevens.edu}
\affiliation{%
  \institution{Stevens Institute of Technology}
  \city{Hoboken}
  \state{NJ}
  \country{USA}}

\renewcommand{\shortauthors}{Trovato et al.}

\begin{abstract}
In the realm of financial analytics, leveraging unstructured data, such as earnings conference calls (ECCs), to forecast stock volatility is a critical challenge that has attracted both academics and investors. While previous studies have used multimodal deep learning-based models to obtain a general view of ECCs for volatility predicting, they often fail to capture detailed, complex information. Our research introduces a novel framework: \textbf{ECC Analyzer}, which utilizes large language models (LLMs) to extract richer, more predictive content from ECCs to aid the model's prediction performance. We use the pre-trained large models to extract textual and audio features from ECCs and implement a hierarchical information extraction strategy to extract more fine-grained information. This strategy first extracts paragraph-level general information by summarizing the text and then extracts fine-grained focus sentences using Retrieval-Augmented Generation (RAG). These features are then fused through multimodal feature fusion to perform volatility prediction. Experimental results demonstrate that our model outperforms traditional analytical benchmarks, confirming the effectiveness of advanced LLM techniques in financial analysis.
\end{abstract}

\ccsdesc[500]{Computing methodologies~Natural language processing}
\ccsdesc[500]{Information systems~Multimedia information systems}

\keywords{Large Language Model, Earnings Conference Call Analysis, Volatility forecasting, Retrieval-Augmented Generation}



\maketitle
\section{Introduction}
Predicting the stock volatility over a certain period is a crucial task in financial analysis, aiding capital market participants in making better investment decisions. As such, developing effective techniques for predicting stock volatility has become increasingly important among academics and the financial industry. Previous studies in economics have shown that stock volatility can be predicted using publicly available information~\cite{dumas2009equilibrium}. Substantial research in finance and computational linguistics has made significant progress in predicting stock volatility from various textual sources, such as company disclosed reports~\cite{kogan2009predicting, hoberg2016text, rekabsaz2017volatility}, the transcripts of earnings conference calls~\cite{kimbrough2005effect, wang2014semiparametric} and social media~\cite{bollen2011twitter, ding2015deep, oliveira2017impact}.

Recently, advancements in multimodal learning have enabled the use of more unstructured multimedia data, such as audio recordings of earning conference calls~\cite{qin2019you, li2020maec}, Merge \& Acquisition calls~\cite{sawhney2021multimodal} and the video of CEO's speech~\cite{rajandran2021interdiscursivity}, for stock volatility prediction. Multimodal learning is particularly valuable in this context as it allows for the integration of diverse data sources, providing richer, more nuanced insights into market dynamics and sentiment. This comprehensive analysis is essential for understanding complex market behaviors~\cite{qin2019you}. Our study focuses on multimodal earning conference calls (ECCs) data for two primary reasons: 1) transcripts and audio recordings are publicly available, and 2) ECCs are often associated with high volatility primarily due to the market reaction to the earnings announcement~\cite{foster1984earnings}. 

Existing multimodal approaches to volatility prediction using ECCs typically extract features from speech and text separately. Subsequently, commonly used Natural Language Processing (NLP) models, such as LSTM or Transformer-based models, are employed to jointly model the extracted speech and text features, ultimately performing volatility prediction~\cite{qin2019you, yang2020html, wang2024ama}. While these methods have shown that the multimodal approach can extract complementary information from multiple modalities, enhancing financial modeling performance and model robustness, several challenges remain unaddressed: firstly, previous work directly feeds features extracted from text and audio into the model, potentially missing important contextual details; secondly, these approaches assign equal importance to all sentences and audio segments which hard to reflect the impact of important information in the ECCs on the prediction.

\begin{figure*}[h]
\centering
  \includegraphics[width=0.95\linewidth]{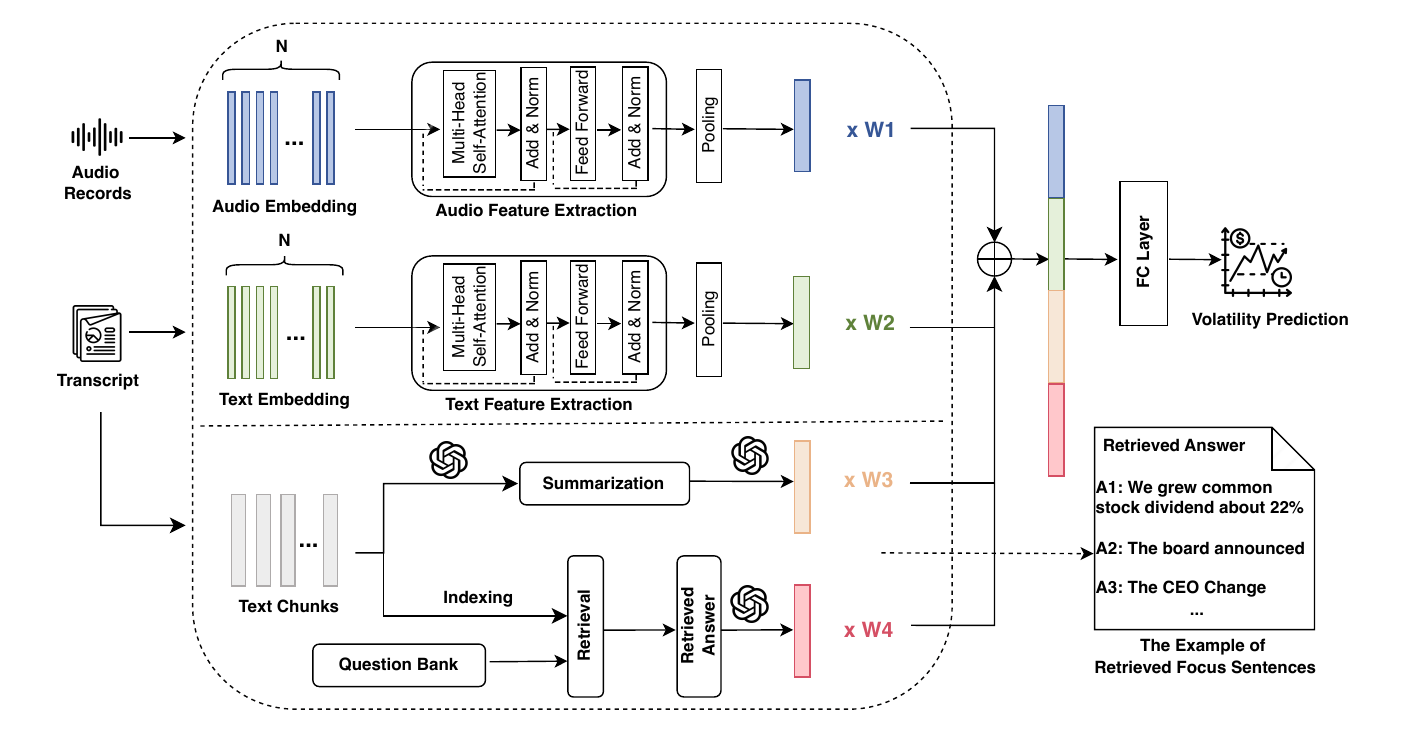} 
  \caption {illustrates the ECC Analyzer Framework. The proposed method accepts multimodal inputs: audio record and transcript. The upper part of the box illustrates the feature extraction process for both the audio and text of the data. We use the pre-trained large models to generate embeddings from the audio and text, followed by a transformer encoder to extract the corresponding features. The lower part of the box represents a deeper analysis of the ECC. Here, the text is divided into chunks, which are then summarized into paragraphs using an LLM. Key sentences are extracted via RAG and converted into text features through text embedding. These text features are then fused with the features extracted from the upper part of the box to make the final volatility prediction.}
  \label{ECC_Analyzer_Framework}
\end{figure*}

In response to the aforementioned limitations and inspired by the significant performance improvements offered by large language models (LLMs) in various downstream NLP tasks, this paper introduces the ECC Analyzer — a novel framework that leverages LLMs for in-depth analysis of ECC data to enhance volatility prediction performance. The proposed ECC Analyzer employs a hierarchical information extraction strategy: 1) advanced pre-trained large models are used to extract embeddings from both audio and text, capturing the overall information content of ECCs.; 2) LLMs summarize ECC transcripts at the paragraph level, distilling important information from the transcript; 3) In collaboration with financial experts, we designed the ``Question Bank" which contains a series of questions about ECC content that are of interest to investors. These questions are used as queries to extract fine-grained sentences containing key information from ECCs using Retrieval-Augmented Generation (RAG). We use these questions as queries and extract fine-grained sentences containing key information from ECC by RAG to improve the prediction performance of the model. Then the summarized text with extracted key focus sentences is transformed into text embedding and combined with the extracted audio and text features for volatility prediction.

Our extensive experiments on the real-world S\&P 500 ECCs dataset demonstrate clear and significant improvements in prediction accuracy attributable to our proposed approach. 
\textbf{We claim that the contribution is two-fold: 1) Our framework innovatively integrates multi-modal information with fine-grained details extracted by Large Language Models (LLMs) to effectively distill and utilize rich information from financial documents, such as ECCs, for prediction tasks. 2) Our pipeline demonstrates a $27.7\%$ reduction in average Mean Squared Error (MSE) compared to the current state-of-the-art (SOTA) model. In detail, The results show that our method achieves substantial performance improvements in forecasting short-term volatility (3-day, 7-day) compared to current SOTA methods. The results for medium-term volatility (15 days) and long-term volatility (30 days) are comparable to the best existing methods.}

\section{Related Work}
\subsection{Stock Volatility Prediction}
Volatility forecasting has long been of interest to researchers due to its practical applications. Conventional forecasting approaches rely on historical stock prices and use continuous and discrete time series models~\cite{kristjanpoller2014volatility, cox1976valuation, heston1993closed, hsu2008constant, franses1996forecasting, kim2018forecasting}. Recently, the use of NLP techniques to analyze unstructured data for predicting stock performance has attracted significant academic attention. A foundational study by~\cite{kogan2009predicting} shows that simple bag-of-words features from annual reports when combined with historical volatility, can outperform models based solely on historical data. Subsequent research, such as that by~\cite{wang2014semiparametric, rekabsaz2017volatility,theil2018word}, proposed various document representation methods to predict stock price volatility. Drawing on multimodal technologies, ~\cite{qin2019you} explored how audio features—such as tone, emotion, and speech rate—enhance stock movement predictions when combined with text analysis. Following by this,~\cite{yang2020html} further extends the idea of using multimodal data to improve risk prediction performance in multi-task learning, and the authors' experiments show that predicting multiple tasks at the same time can help the model further improve prediction performance. \cite{wang2024ama} addressed the reduction of gender bias in ECC predictions through adversarial training. However, the aforementioned studies primarily input ECC data directly into models for prediction without conducting a thorough analysis of the ECC content.

\subsection{Large Language Models in Financial Application}
Numerous studies have explored the applications of LLMs in the financial sector. ~\cite{li2023large} explore how LLMs have been adeptly applied to summarize and abstract complex financial documents such as 10-K, and 10-Q filings. The FinBen~\cite{xie2024finben} provides the benchmark performance of LLM for each task in the financial domain. FinGPT~\cite{yang2023fingpt} and FinMem~\cite{yu2023finmem} explore the usage of LLMs in mining media news for trading recommendations, showcasing the models' ability to discern subtle market indicators and sentiments. In the domain of customer service, the implementation of LLM-powered chatbots is spotlighted for offering context-aware interactions, serving as both assistants and consultants~\cite{lakhani2023enhancing,subagja2023improving,soni2023large}.~\cite{abdaljalil2021exploration,zmandar2021joint} explore the nuanced task of extracting financial and legal items from lengthy text documents, such as financial regulations and comprehensive policy manuals. However, these existing studies predominantly focus on tasks like financial text summarization, question-answering (Q\&A), and stock movement prediction (binary classification), with a notable gap in the application of LLMs for comprehensive stock volatility prediction.

\section{Problem Formulation}
\label{formulation}
Volatility can be expressed as the natural log of the standard deviation of return prices \(r\) in a window of \({\tau}\) days. We calculate the 3, 7, 15, and 30 trading days volatility using the equation:
\begin{equation}
\label{e:vol}
v_{[d-\tau, d]} = \ln\left(\left(\frac{\sum_{i=0}^{\tau}(r_{d-i}-\bar{r})^{2}}{\tau}\right)^{\frac{1}{2}}\right) 
\end{equation}
The notations used across the paper are discussed herein. Let \(s \in S\) denote a stock, \(c \in C\) be an earnings call for stock s. For each stock, there exist multiple earnings calls c that are held periodically. Each call c can be segmented into a set of \(a_{c}^i \in A_{c}\) audio clips, and corresponding \(t_{c}^i \in T_{c}\) text sentences for \(i \in [1,N]\), where is the \(N\) is the maximum number of audio clips in a call. For each stock, there exists a daily return denoted by \(r_i = \ \frac {p_i - p_{i-1}}{p_{i-1}}\), where \(p_i\) is the adjusted close price at the end of the day and \=r is the average return over a period of \({\tau}\) days. We aims to develop a predictive regression function \(f(c) \rightarrow v_{[d-\tau, d]}\)

For the evaluation, following~\cite{qin2019you}, we assess the accuracy of volatility predictions by comparing the predicted values $y_i$ with the labeled volatility values $\hat{y_i}$. We use Mean Squared Error (MSE) as the evaluation metric:
\begin{equation}
MSE = \frac{\sum_{i}(y_i - \hat{y_i})^2}{n}
\end{equation}

This metric quantifies the average squared difference between the predicted and actual volatility values, providing a measure of prediction accuracy. Additionally, for constructing volatility labels, we collected stock price information via the Yahoo Finance API\footnote{https://finance.yahoo.com/} and used equation~\ref{e:vol} to calculate the labels for days 3, 7, 15, and 30.

\section{Our proposed framework}

ECC Analyzer (in Figure~\ref{ECC_Analyzer_Framework}) aims to comprehensively understand the multi-data types present in earnings conference calls, including both text and audio components. In this section, we explain our methodology in 4 parts: (\ref{Audio Encoding}) audio encoding, (\ref{text Encoding}) transcript encoding,(\ref{ECC analysis}) fine-grained information extraction from the transcript by using LLM, and (\ref{model_training}) multi-model fusion and model training. 

\subsection{Audio Encoding} \label{Audio Encoding}
Audio pre-trained models have achieved performing results in various downstream tasks~\cite{pons2019musicnn, cramer2019look, koh2021comparison, wang2021modular}. We aim to leverage advanced audio pre-trained models like Wav2vec2, a transformer-based Large Language Model recognized for its effectiveness in processing raw audio~\cite{baevski2020wav2vec}, to extract audio embeddings. After that, we employ a Multi-Head Self-Attention (MHSA) mechanism to distill specific audio features. This method is vital for integrating these features with other data modalities, facilitating a more detailed and comprehensive analysis.

\begin{figure*}[t]
\centering
  \includegraphics[width=0.95\linewidth]{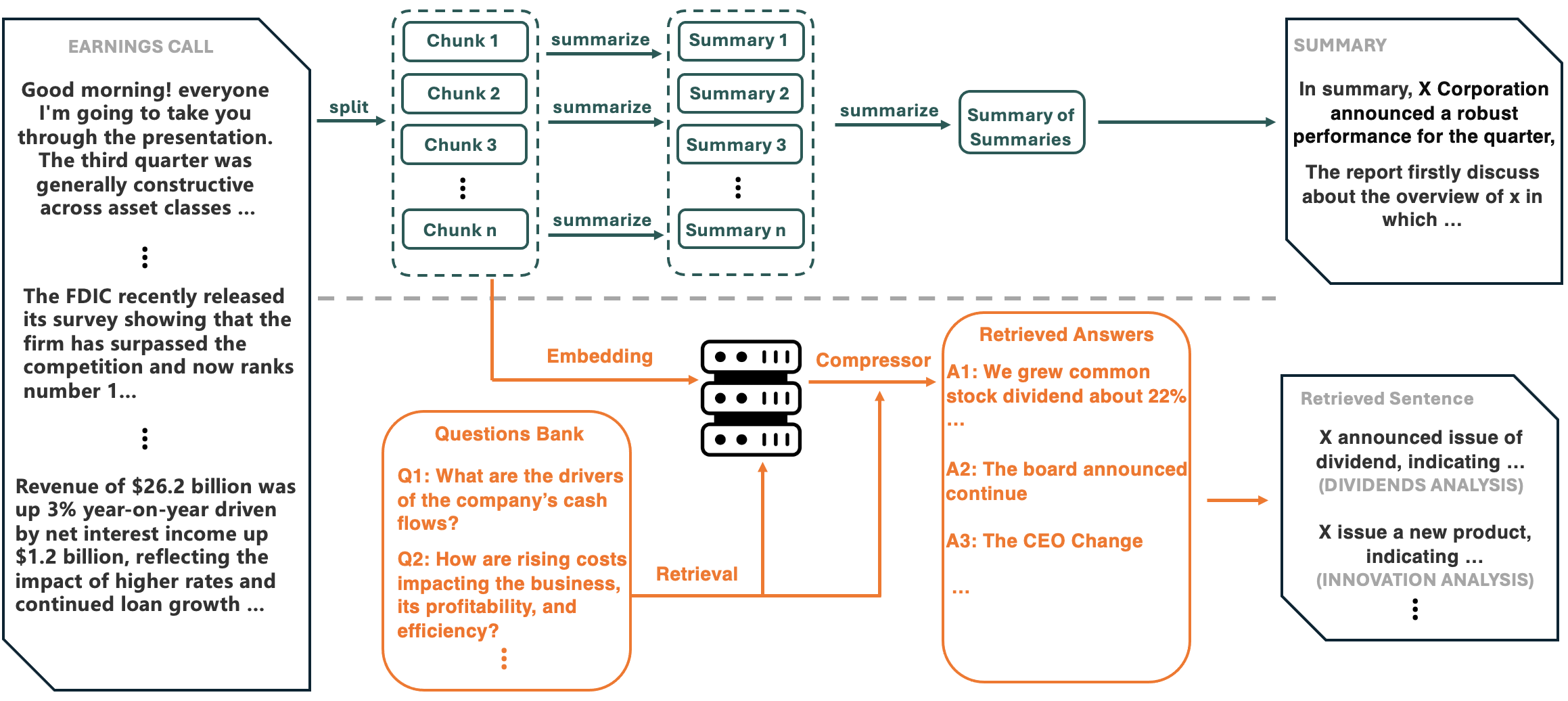} 
  \caption {visualizes the process of fine-grained information extraction from ECC transcript. 
  }
  \label{ECC_Transcript_Analysis}
\end{figure*}

The raw audio input data be represented by \(A_c = \{a_{c}^1, a_{c}^2, \ldots, a_{c}^n \}\) where \(a_c^i\) represents the \(i^{th}\) audio frame in one data sample. Each audio frame will be converted into a vector representation: $e_{ac}^i = \text{Wav2Vec2}(a_c^i)$. Therefore, we obtain the audio embeddings \(E_{ac} = \{e_{ac}^1, e_{ac}^2, \ldots, e_{ac}^n \}\) which have dimensions of 520 $\times$ 512, representing the maximum number of audio files across companies and the transform dimensions for a single audio frame, respectively. Audio files with fewer than 520 frames ($n < 520$) are zero-padded for consistent matrix size.

\(E_{ac}\) are then processed through a MHSA to distill specific audio features. The MHSA includes a multi-head attention block, a norm block, and a two-layer feed-forward network with ReLU activation, forming the basis for all subsequent architectures discussed. In detail, the MHSA calculation process is as follows:
\begin{equation}
\text{Multihead} =  \text{Concat}\text{(head}_1, ...., \text{head}_h)W^o
\end{equation}
\begin{equation}
\text{head}_i =  \text{Attention}(QW_i^Q, KW_i^K, VW_i^V)
\end{equation}

where $Q$ (queries) and $K$ (keys) of dimension $d_k$ and $V$ values of dimension $d_v$. The weights dimensions are: $W_i^Q, W_i^K, W_i^V \in R^{d_{model} \times d_k, d_k, d_v}$ respectively, and $W^o \in R^{d_{v}\times d_{model}}$. The dot product is then calculated for the query with all the keys. The attention scores are normalized using the softmax function:
\begin{equation}
\text{Attention}(Q, K, V) = softmax(\frac{KQ^T}{\sqrt{d_k}})V
\end{equation}

The attention function on a set of queries is calculated simultaneously packed together in a matrix Q. The keys and values are also packed in the matrices K and V respectively. Combining (2)-(4), this results in a matrix:
\begin{equation}
    T_{ac} =  \text{MHSA}(E_{ac})
\end{equation}
where \(T_{ac} = \{t_{ac}^1, t_{ac}^2, \ldots, t_{ac}^n \}\) with size 520 $\times$ 512. \(T_{ac} \) is then subjected to an average pooling layer to produce $T_{a}$, a condensed audio feature vector of size 512.

\subsection{Transcript Encoding} \label{text Encoding}
The transcript encoding process mirrors Audio Encoding, using SimCSE~\cite{gao2021simcse} to extract sentence-level vector representations from earnings conference transcripts. SimCSE is a Siamese neural network architecture that learns to embed pairs of sentences into a shared space where similar sentences are mapped close together and dissimilar sentences are mapped far apart. The raw transcripts are represented as \(T_c\ = \{t_{c}^1, a_{c}^2, \ldots, t_{c}^n \}\), with each sentence \(t_c^i\) represents the \(i^{th}\) transformed into a vector representation: $e_{tc}^i = \text{SimCSE}(t_c^i).$

We obtain the corresponding text embeddings given by \(E_{tc} = \{e_{tc}^1, e_{tc}^2, \ldots, e_{tc}^n \}\) with size 520 $\times$ 768, where 520 is the maximum number of sentences amongst all data samples and 768 is the dimension of the output of SimCSE. Earnings conference calls with less than 520 sentences ($n < 520$) have been zero-padded for uniformity in input matrix size. Same with (1)-(4), the MHSA is applied to \(E_{tc}\) to get \(T_{tc} = \{t_{tc}^1, t_{tc}^2, \ldots, t_{tc}^n \}\) with dimension 520 $\times$ 768. Then,\(T_{tc} \) is subjected to the average pooling layer to produce \(T_{t} \), where  \(T_{t}\) denotes the resultant extracted textual feature of size 768.

\subsection{Fine-grained Information Extraction from the Transcript by Using LLM} \label{ECC analysis}
To obtain deep insights from an Earnings Conference Call on how it might predict future market performance, our approach encompasses two parts: (1) summarize the transcript, and (2) important sentences extracted by using RAG. We show this process in Figure~\ref{ECC_Transcript_Analysis}. We list all Prompts in Appendix~\ref{ssec:prompt}.
\subsubsection{Summarize the transcript} In order to effectively summarize key information in lengthy ECC records, we employ a hierarchical summarization strategy. First, the entire document is divided into chunks, and then we use LLM to summarize each chunk individually. These individual summaries are then further summarized through LLM, resulting in a comprehensive summarization paragraph of the entire document. This two-layer approach ensures that both detailed and aggregate information is captured. In further, we use the OpenAI `text-embedding-3-small' text embedding model to generate embeddings \(T_s\) with size 1024 for summarized paragraph:
\begin{equation}
T_s = \text{Embedding}(summary + chunk\ summaries)
\end{equation}

\subsubsection{Important sentences extracted by using RAG} In this step, we aim to extract the most critical sentences containing relevant information from the entire transcript. To achieve this, we first convert the processed chunks into vector representations using the embedding model. Then, with input from financial experts, we design a list of questions about the ECC, which are stored in a 'Question Bank' as a query list (see Appendix ~\ref{ssec:question}). For each question in the 'Question Bank', we retrieve the relevant chunks from the vectors individually. Next, we use the context compressor, in combination with the LLM chat, to quiz each extracted chunk and query on its relevance. This process filters out only the sentences useful for answering the questions. Finally, we synthesize the query-selected sentences into a coherent prompt, to which the large language model will generate responses as the final answers to the queries. These generated answers are considered as the extracted important sentences. We then concatenate all the generated sentences and feed them into a text embedding model to generate the text embedding \(T_f\) with size 1024:
\begin{equation}
T_f = \text{Embedding}(\text{Concatenated}[selected\ sentences;])
\end{equation}

\subsection{Multimodal Fusion and Prediction} \label{model_training}
Given the model's reliance on several inputs and diverse data types, we identify an effective fusion structure to integrate these features into the training process to ensure a balanced weighting among components.
We use additive interactions to handle the representational fusion of different abstract representations. These operators can be viewed as differentiable building blocks that combine information from several different data streams and can be flexibly inserted into almost any unimodal pipeline~\cite{liang2022foundations}.  Given the audio feature $T_a$, textual feature $T_t$ from the transcript, and $T_s, T_f$ from ECC analyzed text, additive fusion can be seen as learning a new joint representation:
\begin{equation}
    E = w_0 + w_1\cdot T_a + w_2\cdot T_t + w_3\cdot T_s + w_4\cdot T_f + \epsilon
\end{equation}
where $w_1 \in R^{512 \times 512}$, $w_2 \in R^{768 \times 512}$ and $w_3, w_4 \in R^{1024 \times 512}$ are the weights learned for additive fusion, $w_0$ the bias term and $\epsilon$ the error term. $E$ is a vector with 512 as the final feature. This unified feature set $E$ is fed into two fully connected layers to perform the regression task. We train ECC Analyzer by optimizing the MSE loss:
\begin{equation}
\begin{aligned}
    \mathcal{L} &= \mu(\sum_{i}(\hat{y_i}-y_i)^2)
 \end{aligned}
\end{equation}

\section{Experiment}
We describe the datasets used in the experiment and the baseline settings for performance comparisons. Our experiments aim to answer the following research questions (RQs): 
\textbf{RQ1}: Can our proposed pipeline, when integrated with LLMs, significantly improve volatility prediction performance compared to current state-of-the-art (SOTA) approaches?
\textbf{RQ2}: Does the synergy between our model and the LLM result in better performance compared to raw LLM predictions?
\textbf{RQ3}: Does each distinct component derived from ECC contribute to improving volatility prediction accuracy?

\subsection{Dataset}
The dataset utilized in this study is sourced from the publicly available S\&P 500 ECC dataset as constructed by~\cite{qin2019you}. It includes both audio recordings and corresponding text transcripts from the 2017 earnings calls of 500 major companies listed on the S\&P 500 and traded on U.S. stock exchanges. The dataset consists of 572 unique instances where the audio recordings were accurately and closely aligned with the text transcripts. Following the setup by~\cite{qin2019you}, we partitioned the dataset into a training set and a test set with an 8:2 ratio, organized temporally to ensure that the data in the training set precedes those in the test set. This temporal division is crucial for maintaining the integrity of our predictive model, aligning the training process with the principle of using historical data to predict future risks—thus enhancing the accuracy and reliability of our forecasting approach. 

\subsection{Baseline Setup}
We compare our approach to several important baselines including:
\label{subsec:baseline}
\begin{itemize}
    \item \textbf{Classical Method:} We incorporate the GARCH model and its derivatives, as described in ~\cite{franses1996forecasting, kim2018forecasting}. This model is well-recognized for short-term volatility prediction but may not be as effective for forecasting average volatility over longer periods, such as n-day volatility.
    \item \textbf{LSTM~\cite{gers2000learning}}: The Long Short-Term Memory (LSTM) based method is a popular choice for financial time series prediction due to its efficacy in handling sequential data. We use a straightforward LSTM model as a benchmark for volatility prediction.
    \item \textbf{MT-LSTM-ATT~\cite{luong2015multi}:} combines the prediction of average n-day volatility with the forecasting of single-day volatility, employing an attention-enhanced LSTM as the foundational model.
    \item \textbf{HAN (Glove):} uses a Hierarchical Attention Network with dual-layered attention at the word and sentence levels. HAN first gets word embeddings using pre-trained Glove vectors and then processed by a Bi-GRU~\cite{chung2014empirical} encoder, while another Bi-GRU encoder simultaneously forms a sentence-level representation of each document. The resulting document representation is input into a regression layer to produce predictions.
    
    \item \textbf{MRDM~\cite{qin2019you}:} The MRDM model first introduced a multi-modal deep regression approach to fuse the GloVe embeddings and hand-crafted acoustic features for volatility prediction tasks. 
    \item \textbf{HTML~\cite{yang2020html}:} This work presented a state-of-the-art model that employs WWM-BERT for text token encoding. Similar to MDRM, HTML also leverages the same audio features. These unimodal features are then combined and processed through a sentence-level transformer, resulting in multimodal representations for each call.
     \item \textbf{AMA-LSTM~\cite{wang2024ama}:} The paper uses a strategy of adversarial training to minimize the effect of speaker gender in earnings conference calls.
    \item \textbf{GPT-4-turbo-2024-04-09:} We assessed the capability of LLMs in directly predicting volatility performance from ECCs. The model was set to generate a response with a zero temperature setting to ensure deterministic output.
\end{itemize}

\subsection{Implementation Details}
In the experiment, all interactions with LLMs were conducted using “GPT-4 Turbo-2024-04-09”. We set the temperature parameter to 0 to ensure that the LLMs produce the most predictable responses, thereby maintaining consistency in our experiments.

For the overall training of the ECC Analyer framework, we developed the code using PyTorch. Each Multi-Head Self-Attention layer in the model comprises 6 layers and 8 individual heads in each layer. The training process utilized batch sizes $b \in \{2, 4, 8, 16\}$. We use a grid search to determine the optimal parameters and select the learning rate $\lambda$ for Adam optimizer among $\{1e-3, 1e-5, 1e-6, 1e-7\}$. 

\subsection{Overall Results Analysis (RQ1 \& RQ2)}
\begin{table}[h]
\centering
\caption{Performance results on our proposed framework ECC Analyzer with different baseline methods.}
\scalebox{1.0}{
\begin{tabular}{lccccc}
\toprule
\textbf{Model}   & \textbf{$\overline{MSE}$}         & \textbf{$MSE_3$} & \textbf{$MSE_7$} & \textbf{$MSE_{15}$} & \textbf{$MSE_{30}$} \\ \midrule
Classical Method & 0.713 &1.710   & 0.526  & 0.330   & 0.284        \\
LSTM             & 0.746 &1.970   & 0.459  & 0.320   & 0.235           \\
MT-LSTM-ATT      & 0.739 &1.983   & 0.435  & 0.304   & 0.233            \\
HAN (Glove)              & 0.598 &1.426   & 0.461  & 0.308   & 0.198        \\
MRDM             & 0.577 &1.371   & 0.420  & 0.300   & 0.217     \\
HTML             & 0.401 &0.845   & 0.349  & 0.251   & \textbf{0.158}         \\
AMA-LSTM             & / &0.680   & 0.360  & \textbf{0.230}   & /         \\
GPT-4-Turbo   & 2.198 & 3.187             & 5.059             & 7.959              & 11.824            \\ \midrule
\textbf{ECC Analyzer} & \textbf{0.314} & \textbf{0.553} & \textbf{0.306} & 0.237 & \textbf{0.158}  \\ 
\bottomrule
\end{tabular}}
\label{tab:overall_performance}
\end{table}

\begin{table*}[h]
  \centering
  \caption{Performance results of ablation study. We designed the ablation study as follows: 1) Audio+Text: uses raw audio and text data from ECCs; 2)Audio+Text+$E_{os}$: adds an overall ECC summary  generated by LLMs; 3)Audio+Text+$E_{cs}$: adds the chucks summary generated by LLMs; 4) Audio+Text+$E_{os}+E_{cs}$: integrates both overall and chunk summaries for the ECC; 5) Audio+Text+$E_{fo}$: combines raw data with focused analytical results; 6) $E_{os}+E_{cs}+E_{fo}$: merges all LLM analyses for prediction without raw data; 6) Audio+Text+$E_{os}+E_{cs}+E_{fo}$: combines all data and analyses for enhanced prediction.}
  \scalebox{1}{
    \begin{tabular}{lccccc}
    \toprule
    \textbf{Module} & $\overline{MSE}$   & $MSE_3$ & $MSE_7$ & $MSE_{15}$ & $MSE_{30}$ \\
    \midrule
    Audio+Text & 0.373 & 0.645 & 0.362 & 0.280  & 0.204 \\
    Audio+Text+$E_{os}$ & 0.373     & 0.638     & 0.380     & 0.276     & 0.201     \\
    Audio+Text+$E_{cs}$ & 0.375     & 0.640     & 0.385     & 0.275     & 0.201     \\
    Audio+Text+$E_{os}+E_{cs}$ & 0.357     & 0.627     & 0.335     & 0.267     & 0.199    \\
    Audio+Text+$E_{fo}$ & 0.324 & 0.579 & 0.323 & 0.230  & 0.165 \\
    $E_{os}+E_{cs}+E_{fo}$ & 0.343 & 0.601 & 0.344 & 0.247 & 0.179 \\
    Audio+Text+$E_{os}+E_{cs}+E_{fo}$ & \textbf{0.314} & \textbf{0.553} & \textbf{0.306} & \textbf{0.237}  & \textbf{0.158}  \\
    \bottomrule
    \end{tabular}}%
  \label{tab:addlabel}%
\end{table*}%

Table~\ref{tab:overall_performance} shows the performance of various methods in predicting volatility performance. Notably, the ECC Analyzer framework excels, especially in short-term forecasts (day 3 and day 7), with the lowest Mean Squared Error (MSE) values of 0.553 and 0.306, respectively. Its long-term prediction performance is comparable to the state-of-the-art method, HTML. However, the ECC Analyzer performs slightly lower than AMA-LSTM in predicting medium-term (day 15) volatility. Nevertheless, ECC Analyzer's overall average MSE achieves superior performance. These encouraging experimental results illustrate that extracting fine-grained information from ECC data using LLMs can significantly enhance the model's volatility prediction accuracy. In particular, the extracted fine-grained information features are especially beneficial for improving short-term forecasting performance, which is crucial for investors. In summary, addressing \textbf{RQ1}, the prediction performance of our proposed method surpasses that of the current SOTA approaches.

In response to \textbf{RQ2}, directly applying LLMs to volatility prediction proves largely ineffective, akin to random guessing. These results also raise concerns about the potential misuse of LLMs, considering their impact on public safety. If investors use LLMs inappropriately for numerical outputs, they may increase financial risk. This indicates that LLMs are more effective as tools to enhance investors' understanding of a company's financial health rather than direct predictors of financial metrics.

\subsection{Ablation Study (RQ3)}
In our research, we conducted an ablation study to assess how different combinations of ECC analysis results impact our model's performance. This systematic comparison helped us identify the individual contributions of each component. Each component is represented as follows: \textbf{$E_{os}$ represents the embedding of the overall ECC summary; $E_{cs}$ represents the embedding of the summary for chunks; $E_{cs}$ represents the embedding of the extracted fine-grained important sentence.}
 
 According to Table~\ref{tab:addlabel}, we can find that the audio and text features extracted by advanced large language models significantly improved short-term prediction accuracy compared to previous methods. Furthermore, incorporating summaries of the data slightly enhanced performance, but more notable improvements were observed when we added analysis of specific focus points. This indicates that our model effectively isolates and utilizes the most relevant information for predicting stock movements. Our best analytical results come from integrating the full spectrum of data and analytical outputs, underscoring the value of each component in our model. We also obtain good predictive results using only analytics derived from LLMs, affirming the response to our \textbf{RQ3}: comprehensive analysis indeed enhances the predictive capability for stock volatility prediction performance.

Our findings also suggest that while earnings calls are information-rich, including every detail in the analysis can be counterproductive and may cloud essential insights. It is therefore critical to pinpoint and concentrate on the most predictive elements of the data, filtering out less relevant information to optimize the analysis process for stock performance prediction.

\section{Conclusion}
In this paper, we propose the ECC Analyzer, a novel framework that leverages large language models (LLMs) for in-depth analysis of ECC data to enhance volatility prediction performance. The ECC Analyzer extracts information from ECCs at multiple levels to assist in volatility prediction. Our experiments demonstrate that our proposed method improves the overall average Mean Squared Error (MSE). Specifically, we achieve better results in short-term forecasting, with medium-term and long-term forecasts also on par with SOTA methods. Additionally, we analyze the role of each component in the proposed framework for the forecasting process through ablation experiments, highlighting the contribution of each element to the overall performance.


\bibliographystyle{ACM-Reference-Format}
\bibliography{sample-base}


\begin{thebibliography}{42}


\ifx \showCODEN    \undefined \def \showCODEN     #1{\unskip}     \fi
\ifx \showDOI      \undefined \def \showDOI       #1{#1}\fi
\ifx \showISBNx    \undefined \def \showISBNx     #1{\unskip}     \fi
\ifx \showISBNxiii \undefined \def \showISBNxiii  #1{\unskip}     \fi
\ifx \showISSN     \undefined \def \showISSN      #1{\unskip}     \fi
\ifx \showLCCN     \undefined \def \showLCCN      #1{\unskip}     \fi
\ifx \shownote     \undefined \def \shownote      #1{#1}          \fi
\ifx \showarticletitle \undefined \def \showarticletitle #1{#1}   \fi
\ifx \showURL      \undefined \def \showURL       {\relax}        \fi
\providecommand\bibfield[2]{#2}
\providecommand\bibinfo[2]{#2}
\providecommand\natexlab[1]{#1}
\providecommand\showeprint[2][]{arXiv:#2}

\bibitem[Abdaljalil and Bouamor(2021)]%
        {abdaljalil2021exploration}
\bibfield{author}{\bibinfo{person}{Samir Abdaljalil} {and} \bibinfo{person}{Houda Bouamor}.} \bibinfo{year}{2021}\natexlab{}.
\newblock \showarticletitle{An exploration of automatic text summarization of financial reports}. In \bibinfo{booktitle}{\emph{Proceedings of the Third Workshop on Financial Technology and Natural Language Processing}}. \bibinfo{pages}{1--7}.
\newblock


\bibitem[Baevski et~al\mbox{.}(2020)]%
        {baevski2020wav2vec}
\bibfield{author}{\bibinfo{person}{Alexei Baevski}, \bibinfo{person}{Yuhao Zhou}, \bibinfo{person}{Abdelrahman Mohamed}, {and} \bibinfo{person}{Michael Auli}.} \bibinfo{year}{2020}\natexlab{}.
\newblock \showarticletitle{wav2vec 2.0: A framework for self-supervised learning of speech representations}.
\newblock \bibinfo{journal}{\emph{Advances in neural information processing systems}}  \bibinfo{volume}{33} (\bibinfo{year}{2020}), \bibinfo{pages}{12449--12460}.
\newblock


\bibitem[Bollen et~al\mbox{.}(2011)]%
        {bollen2011twitter}
\bibfield{author}{\bibinfo{person}{Johan Bollen}, \bibinfo{person}{Huina Mao}, {and} \bibinfo{person}{Xiaojun Zeng}.} \bibinfo{year}{2011}\natexlab{}.
\newblock \showarticletitle{Twitter mood predicts the stock market}.
\newblock \bibinfo{journal}{\emph{Journal of computational science}} \bibinfo{volume}{2}, \bibinfo{number}{1} (\bibinfo{year}{2011}), \bibinfo{pages}{1--8}.
\newblock


\bibitem[Chung et~al\mbox{.}(2014)]%
        {chung2014empirical}
\bibfield{author}{\bibinfo{person}{Junyoung Chung}, \bibinfo{person}{Caglar Gulcehre}, \bibinfo{person}{KyungHyun Cho}, {and} \bibinfo{person}{Yoshua Bengio}.} \bibinfo{year}{2014}\natexlab{}.
\newblock \showarticletitle{Empirical evaluation of gated recurrent neural networks on sequence modeling}.
\newblock \bibinfo{journal}{\emph{arXiv preprint arXiv:1412.3555}} (\bibinfo{year}{2014}).
\newblock


\bibitem[Cox and Ross(1976)]%
        {cox1976valuation}
\bibfield{author}{\bibinfo{person}{John~C Cox} {and} \bibinfo{person}{Stephen~A Ross}.} \bibinfo{year}{1976}\natexlab{}.
\newblock \showarticletitle{The valuation of options for alternative stochastic processes}.
\newblock \bibinfo{journal}{\emph{Journal of financial economics}} \bibinfo{volume}{3}, \bibinfo{number}{1-2} (\bibinfo{year}{1976}), \bibinfo{pages}{145--166}.
\newblock


\bibitem[Cramer et~al\mbox{.}(2019)]%
        {cramer2019look}
\bibfield{author}{\bibinfo{person}{Aurora~Linh Cramer}, \bibinfo{person}{Ho-Hsiang Wu}, \bibinfo{person}{Justin Salamon}, {and} \bibinfo{person}{Juan~Pablo Bello}.} \bibinfo{year}{2019}\natexlab{}.
\newblock \showarticletitle{Look, listen, and learn more: Design choices for deep audio embeddings}. In \bibinfo{booktitle}{\emph{ICASSP 2019-2019 IEEE International Conference on Acoustics, Speech and Signal Processing (ICASSP)}}. IEEE, \bibinfo{pages}{3852--3856}.
\newblock


\bibitem[Ding et~al\mbox{.}(2015)]%
        {ding2015deep}
\bibfield{author}{\bibinfo{person}{Xiao Ding}, \bibinfo{person}{Yue Zhang}, \bibinfo{person}{Ting Liu}, {and} \bibinfo{person}{Junwen Duan}.} \bibinfo{year}{2015}\natexlab{}.
\newblock \showarticletitle{Deep learning for event-driven stock prediction}. In \bibinfo{booktitle}{\emph{Twenty-fourth international joint conference on artificial intelligence}}.
\newblock


\bibitem[Dumas et~al\mbox{.}(2009)]%
        {dumas2009equilibrium}
\bibfield{author}{\bibinfo{person}{Bernard Dumas}, \bibinfo{person}{Alexander Kurshev}, {and} \bibinfo{person}{Raman Uppal}.} \bibinfo{year}{2009}\natexlab{}.
\newblock \showarticletitle{Equilibrium portfolio strategies in the presence of sentiment risk and excess volatility}.
\newblock \bibinfo{journal}{\emph{The Journal of Finance}} \bibinfo{volume}{64}, \bibinfo{number}{2} (\bibinfo{year}{2009}), \bibinfo{pages}{579--629}.
\newblock


\bibitem[Foster et~al\mbox{.}(1984)]%
        {foster1984earnings}
\bibfield{author}{\bibinfo{person}{George Foster}, \bibinfo{person}{Chris Olsen}, {and} \bibinfo{person}{Terry Shevlin}.} \bibinfo{year}{1984}\natexlab{}.
\newblock \showarticletitle{Earnings releases, anomalies, and the behavior of security returns}.
\newblock \bibinfo{journal}{\emph{Accounting Review}} (\bibinfo{year}{1984}), \bibinfo{pages}{574--603}.
\newblock


\bibitem[Franses and Van~Dijk(1996)]%
        {franses1996forecasting}
\bibfield{author}{\bibinfo{person}{Philip~Hans Franses} {and} \bibinfo{person}{Dick Van~Dijk}.} \bibinfo{year}{1996}\natexlab{}.
\newblock \showarticletitle{Forecasting stock market volatility using (non-linear) Garch models}.
\newblock \bibinfo{journal}{\emph{Journal of forecasting}} \bibinfo{volume}{15}, \bibinfo{number}{3} (\bibinfo{year}{1996}), \bibinfo{pages}{229--235}.
\newblock


\bibitem[Gao et~al\mbox{.}(2021)]%
        {gao2021simcse}
\bibfield{author}{\bibinfo{person}{Tianyu Gao}, \bibinfo{person}{Xingcheng Yao}, {and} \bibinfo{person}{Danqi Chen}.} \bibinfo{year}{2021}\natexlab{}.
\newblock \showarticletitle{Simcse: Simple contrastive learning of sentence embeddings}.
\newblock \bibinfo{journal}{\emph{arXiv preprint arXiv:2104.08821}} (\bibinfo{year}{2021}).
\newblock


\bibitem[Gers et~al\mbox{.}(2000)]%
        {gers2000learning}
\bibfield{author}{\bibinfo{person}{Felix~A Gers}, \bibinfo{person}{J{\"u}rgen Schmidhuber}, {and} \bibinfo{person}{Fred Cummins}.} \bibinfo{year}{2000}\natexlab{}.
\newblock \showarticletitle{Learning to forget: Continual prediction with LSTM}.
\newblock \bibinfo{journal}{\emph{Neural computation}} \bibinfo{volume}{12}, \bibinfo{number}{10} (\bibinfo{year}{2000}), \bibinfo{pages}{2451--2471}.
\newblock


\bibitem[Heston(1993)]%
        {heston1993closed}
\bibfield{author}{\bibinfo{person}{Steven~L Heston}.} \bibinfo{year}{1993}\natexlab{}.
\newblock \showarticletitle{A closed-form solution for options with stochastic volatility with applications to bond and currency options}.
\newblock \bibinfo{journal}{\emph{The review of financial studies}} \bibinfo{volume}{6}, \bibinfo{number}{2} (\bibinfo{year}{1993}), \bibinfo{pages}{327--343}.
\newblock


\bibitem[Hoberg and Phillips(2016)]%
        {hoberg2016text}
\bibfield{author}{\bibinfo{person}{Gerard Hoberg} {and} \bibinfo{person}{Gordon Phillips}.} \bibinfo{year}{2016}\natexlab{}.
\newblock \showarticletitle{Text-based network industries and endogenous product differentiation}.
\newblock \bibinfo{journal}{\emph{Journal of political economy}} \bibinfo{volume}{124}, \bibinfo{number}{5} (\bibinfo{year}{2016}), \bibinfo{pages}{1423--1465}.
\newblock


\bibitem[Hsu et~al\mbox{.}(2008)]%
        {hsu2008constant}
\bibfield{author}{\bibinfo{person}{Ying-Lin Hsu}, \bibinfo{person}{TI Lin}, {and} \bibinfo{person}{CF Lee}.} \bibinfo{year}{2008}\natexlab{}.
\newblock \showarticletitle{Constant elasticity of variance (CEV) option pricing model: Integration and detailed derivation}.
\newblock \bibinfo{journal}{\emph{Mathematics and Computers in Simulation}} \bibinfo{volume}{79}, \bibinfo{number}{1} (\bibinfo{year}{2008}), \bibinfo{pages}{60--71}.
\newblock


\bibitem[Kim and Won(2018)]%
        {kim2018forecasting}
\bibfield{author}{\bibinfo{person}{Ha~Young Kim} {and} \bibinfo{person}{Chang~Hyun Won}.} \bibinfo{year}{2018}\natexlab{}.
\newblock \showarticletitle{Forecasting the volatility of stock price index: A hybrid model integrating LSTM with multiple GARCH-type models}.
\newblock \bibinfo{journal}{\emph{Expert Systems with Applications}}  \bibinfo{volume}{103} (\bibinfo{year}{2018}), \bibinfo{pages}{25--37}.
\newblock


\bibitem[Kimbrough(2005)]%
        {kimbrough2005effect}
\bibfield{author}{\bibinfo{person}{Michael~D Kimbrough}.} \bibinfo{year}{2005}\natexlab{}.
\newblock \showarticletitle{The effect of conference calls on analyst and market underreaction to earnings announcements}.
\newblock \bibinfo{journal}{\emph{The Accounting Review}} \bibinfo{volume}{80}, \bibinfo{number}{1} (\bibinfo{year}{2005}), \bibinfo{pages}{189--219}.
\newblock


\bibitem[Kogan et~al\mbox{.}(2009)]%
        {kogan2009predicting}
\bibfield{author}{\bibinfo{person}{Shimon Kogan}, \bibinfo{person}{Dimitry Levin}, \bibinfo{person}{Bryan~R Routledge}, \bibinfo{person}{Jacob~S Sagi}, {and} \bibinfo{person}{Noah~A Smith}.} \bibinfo{year}{2009}\natexlab{}.
\newblock \showarticletitle{Predicting risk from financial reports with regression}. In \bibinfo{booktitle}{\emph{Proceedings of human language technologies: the 2009 annual conference of the North American Chapter of the Association for Computational Linguistics}}. \bibinfo{pages}{272--280}.
\newblock


\bibitem[Koh and Dubnov(2021)]%
        {koh2021comparison}
\bibfield{author}{\bibinfo{person}{Eunjeong Koh} {and} \bibinfo{person}{Shlomo Dubnov}.} \bibinfo{year}{2021}\natexlab{}.
\newblock \showarticletitle{Comparison and analysis of deep audio embeddings for music emotion recognition}.
\newblock \bibinfo{journal}{\emph{arXiv preprint arXiv:2104.06517}} (\bibinfo{year}{2021}).
\newblock


\bibitem[Kristjanpoller et~al\mbox{.}(2014)]%
        {kristjanpoller2014volatility}
\bibfield{author}{\bibinfo{person}{Werner Kristjanpoller}, \bibinfo{person}{Anton Fadic}, {and} \bibinfo{person}{Marcel~C Minutolo}.} \bibinfo{year}{2014}\natexlab{}.
\newblock \showarticletitle{Volatility forecast using hybrid neural network models}.
\newblock \bibinfo{journal}{\emph{Expert Systems with Applications}} \bibinfo{volume}{41}, \bibinfo{number}{5} (\bibinfo{year}{2014}), \bibinfo{pages}{2437--2442}.
\newblock


\bibitem[Lakhani(2023)]%
        {lakhani2023enhancing}
\bibfield{author}{\bibinfo{person}{Akbar Lakhani}.} \bibinfo{year}{2023}\natexlab{}.
\newblock \showarticletitle{Enhancing Customer Service with ChatGPT Transforming the Way Businesses Interact with Customers}.
\newblock  (\bibinfo{year}{2023}).
\newblock


\bibitem[Li et~al\mbox{.}(2020)]%
        {li2020maec}
\bibfield{author}{\bibinfo{person}{Jiazheng Li}, \bibinfo{person}{Linyi Yang}, \bibinfo{person}{Barry Smyth}, {and} \bibinfo{person}{Ruihai Dong}.} \bibinfo{year}{2020}\natexlab{}.
\newblock \showarticletitle{Maec: A multimodal aligned earnings conference call dataset for financial risk prediction}. In \bibinfo{booktitle}{\emph{Proceedings of the 29th ACM International Conference on Information \& Knowledge Management}}. \bibinfo{pages}{3063--3070}.
\newblock


\bibitem[Li et~al\mbox{.}(2023)]%
        {li2023large}
\bibfield{author}{\bibinfo{person}{Yinheng Li}, \bibinfo{person}{Shaofei Wang}, \bibinfo{person}{Han Ding}, {and} \bibinfo{person}{Hang Chen}.} \bibinfo{year}{2023}\natexlab{}.
\newblock \showarticletitle{Large language models in finance: A survey}. In \bibinfo{booktitle}{\emph{Proceedings of the Fourth ACM International Conference on AI in Finance}}. \bibinfo{pages}{374--382}.
\newblock


\bibitem[Liang et~al\mbox{.}(2022)]%
        {liang2022foundations}
\bibfield{author}{\bibinfo{person}{Paul~Pu Liang}, \bibinfo{person}{Amir Zadeh}, {and} \bibinfo{person}{Louis-Philippe Morency}.} \bibinfo{year}{2022}\natexlab{}.
\newblock \showarticletitle{Foundations and Trends in Multimodal Machine Learning: Principles, Challenges, and Open Questions}.
\newblock \bibinfo{journal}{\emph{arXiv preprint arXiv:2209.03430}} (\bibinfo{year}{2022}).
\newblock


\bibitem[Luong et~al\mbox{.}(2015)]%
        {luong2015multi}
\bibfield{author}{\bibinfo{person}{Minh-Thang Luong}, \bibinfo{person}{Quoc~V Le}, \bibinfo{person}{Ilya Sutskever}, \bibinfo{person}{Oriol Vinyals}, {and} \bibinfo{person}{Lukasz Kaiser}.} \bibinfo{year}{2015}\natexlab{}.
\newblock \showarticletitle{Multi-task sequence to sequence learning}.
\newblock \bibinfo{journal}{\emph{arXiv preprint arXiv:1511.06114}} (\bibinfo{year}{2015}).
\newblock


\bibitem[Oliveira et~al\mbox{.}(2017)]%
        {oliveira2017impact}
\bibfield{author}{\bibinfo{person}{Nuno Oliveira}, \bibinfo{person}{Paulo Cortez}, {and} \bibinfo{person}{Nelson Areal}.} \bibinfo{year}{2017}\natexlab{}.
\newblock \showarticletitle{The impact of microblogging data for stock market prediction: Using Twitter to predict returns, volatility, trading volume and survey sentiment indices}.
\newblock \bibinfo{journal}{\emph{Expert Systems with applications}}  \bibinfo{volume}{73} (\bibinfo{year}{2017}), \bibinfo{pages}{125--144}.
\newblock


\bibitem[Pons and Serra(2019)]%
        {pons2019musicnn}
\bibfield{author}{\bibinfo{person}{Jordi Pons} {and} \bibinfo{person}{Xavier Serra}.} \bibinfo{year}{2019}\natexlab{}.
\newblock \showarticletitle{musicnn: Pre-trained convolutional neural networks for music audio tagging}.
\newblock \bibinfo{journal}{\emph{arXiv preprint arXiv:1909.06654}} (\bibinfo{year}{2019}).
\newblock


\bibitem[Qin and Yang(2019)]%
        {qin2019you}
\bibfield{author}{\bibinfo{person}{Yu Qin} {and} \bibinfo{person}{Yi Yang}.} \bibinfo{year}{2019}\natexlab{}.
\newblock \showarticletitle{What you say and how you say it matters: Predicting stock volatility using verbal and vocal cues}. In \bibinfo{booktitle}{\emph{Proceedings of the 57th Annual Meeting of the Association for Computational Linguistics}}. \bibinfo{pages}{390--401}.
\newblock


\bibitem[Rajandran(2021)]%
        {rajandran2021interdiscursivity}
\bibfield{author}{\bibinfo{person}{Kumaran Rajandran}.} \bibinfo{year}{2021}\natexlab{}.
\newblock \showarticletitle{Interdiscursivity in corporate financial communication: an analysis of earnings videos}.
\newblock \bibinfo{journal}{\emph{Corporate Communications: An International Journal}} \bibinfo{volume}{26}, \bibinfo{number}{2} (\bibinfo{year}{2021}), \bibinfo{pages}{328--347}.
\newblock


\bibitem[Rekabsaz et~al\mbox{.}(2017)]%
        {rekabsaz2017volatility}
\bibfield{author}{\bibinfo{person}{Navid Rekabsaz}, \bibinfo{person}{Mihai Lupu}, \bibinfo{person}{Artem Baklanov}, \bibinfo{person}{Allan Hanbury}, \bibinfo{person}{Alexander D{\"u}r}, {and} \bibinfo{person}{Linda Anderson}.} \bibinfo{year}{2017}\natexlab{}.
\newblock \showarticletitle{Volatility prediction using financial disclosures sentiments with word embedding-based IR models}.
\newblock \bibinfo{journal}{\emph{arXiv preprint arXiv:1702.01978}} (\bibinfo{year}{2017}).
\newblock


\bibitem[Sawhney et~al\mbox{.}(2021)]%
        {sawhney2021multimodal}
\bibfield{author}{\bibinfo{person}{Ramit Sawhney}, \bibinfo{person}{Mihir Goyal}, \bibinfo{person}{Prakhar Goel}, \bibinfo{person}{Puneet Mathur}, {and} \bibinfo{person}{Rajiv Shah}.} \bibinfo{year}{2021}\natexlab{}.
\newblock \showarticletitle{Multimodal multi-speaker merger \& acquisition financial modeling: A new task, dataset, and neural baselines}. In \bibinfo{booktitle}{\emph{Proceedings of the 59th Annual Meeting of the Association for Computational Linguistics and the 11th International Joint Conference on Natural Language Processing (Volume 1: Long Papers)}}. \bibinfo{pages}{6751--6762}.
\newblock


\bibitem[Soni(2023)]%
        {soni2023large}
\bibfield{author}{\bibinfo{person}{Vishvesh Soni}.} \bibinfo{year}{2023}\natexlab{}.
\newblock \showarticletitle{Large language models for enhancing customer lifecycle management}.
\newblock \bibinfo{journal}{\emph{Journal of Empirical Social Science Studies}} \bibinfo{volume}{7}, \bibinfo{number}{1} (\bibinfo{year}{2023}), \bibinfo{pages}{67--89}.
\newblock


\bibitem[Subagja et~al\mbox{.}(2023)]%
        {subagja2023improving}
\bibfield{author}{\bibinfo{person}{Agus~Dedi Subagja}, \bibinfo{person}{Abu Muna~Almaududi Ausat}, \bibinfo{person}{Ade~Risna Sari}, \bibinfo{person}{M~Indre Wanof}, {and} \bibinfo{person}{Suherlan Suherlan}.} \bibinfo{year}{2023}\natexlab{}.
\newblock \showarticletitle{Improving customer service quality in MSMEs through the use of ChatGPT}.
\newblock \bibinfo{journal}{\emph{Jurnal Minfo Polgan}} \bibinfo{volume}{12}, \bibinfo{number}{1} (\bibinfo{year}{2023}), \bibinfo{pages}{380--386}.
\newblock


\bibitem[Theil et~al\mbox{.}(2018)]%
        {theil2018word}
\bibfield{author}{\bibinfo{person}{Christoph~Kilian Theil}, \bibinfo{person}{Sanja {\v{S}}tajner}, {and} \bibinfo{person}{Heiner Stuckenschmidt}.} \bibinfo{year}{2018}\natexlab{}.
\newblock \showarticletitle{Word embeddings-based uncertainty detection in financial disclosures}. In \bibinfo{booktitle}{\emph{Proceedings of the first workshop on economics and natural language processing}}. \bibinfo{pages}{32--37}.
\newblock


\bibitem[Wang et~al\mbox{.}(2021)]%
        {wang2021modular}
\bibfield{author}{\bibinfo{person}{Ning Wang}, \bibinfo{person}{Yupeng Cao}, \bibinfo{person}{Shuai Hao}, \bibinfo{person}{Zongru Shao}, {and} \bibinfo{person}{KP Subbalakshmi}.} \bibinfo{year}{2021}\natexlab{}.
\newblock \showarticletitle{Modular Multi-Modal Attention Network for Alzheimer's Disease Detection Using Patient Audio and Language Data.}. In \bibinfo{booktitle}{\emph{Interspeech}}. \bibinfo{pages}{3835--3839}.
\newblock


\bibitem[Wang et~al\mbox{.}(2024)]%
        {wang2024ama}
\bibfield{author}{\bibinfo{person}{Shengkun Wang}, \bibinfo{person}{Taoran Ji}, \bibinfo{person}{Jianfeng He}, \bibinfo{person}{Mariam Almutairi}, \bibinfo{person}{Dan Wang}, \bibinfo{person}{Linhan Wang}, \bibinfo{person}{Min Zhang}, {and} \bibinfo{person}{Chang-Tien Lu}.} \bibinfo{year}{2024}\natexlab{}.
\newblock \showarticletitle{AMA-LSTM: Pioneering Robust and Fair Financial Audio Analysis for Stock Volatility Prediction}.
\newblock \bibinfo{journal}{\emph{arXiv preprint arXiv:2407.18324}} (\bibinfo{year}{2024}).
\newblock


\bibitem[Wang and Hua(2014)]%
        {wang2014semiparametric}
\bibfield{author}{\bibinfo{person}{William~Yang Wang} {and} \bibinfo{person}{Zhenhao Hua}.} \bibinfo{year}{2014}\natexlab{}.
\newblock \showarticletitle{A semiparametric gaussian copula regression model for predicting financial risks from earnings calls}. In \bibinfo{booktitle}{\emph{Proceedings of the 52nd Annual Meeting of the Association for Computational Linguistics (Volume 1: Long Papers)}}. \bibinfo{pages}{1155--1165}.
\newblock


\bibitem[Xie et~al\mbox{.}(2024)]%
        {xie2024finben}
\bibfield{author}{\bibinfo{person}{Qianqian Xie}, \bibinfo{person}{Weiguang Han}, \bibinfo{person}{Zhengyu Chen}, \bibinfo{person}{Ruoyu Xiang}, \bibinfo{person}{Xiao Zhang}, \bibinfo{person}{Yueru He}, \bibinfo{person}{Mengxi Xiao}, \bibinfo{person}{Dong Li}, \bibinfo{person}{Yongfu Dai}, \bibinfo{person}{Duanyu Feng}, {et~al\mbox{.}}} \bibinfo{year}{2024}\natexlab{}.
\newblock \showarticletitle{The finben: An holistic financial benchmark for large language models}.
\newblock \bibinfo{journal}{\emph{arXiv preprint arXiv:2402.12659}} (\bibinfo{year}{2024}).
\newblock


\bibitem[Yang et~al\mbox{.}(2023)]%
        {yang2023fingpt}
\bibfield{author}{\bibinfo{person}{Hongyang Yang}, \bibinfo{person}{Xiao-Yang Liu}, {and} \bibinfo{person}{Christina~Dan Wang}.} \bibinfo{year}{2023}\natexlab{}.
\newblock \showarticletitle{Fingpt: Open-source financial large language models}.
\newblock \bibinfo{journal}{\emph{arXiv preprint arXiv:2306.06031}} (\bibinfo{year}{2023}).
\newblock


\bibitem[Yang et~al\mbox{.}(2020)]%
        {yang2020html}
\bibfield{author}{\bibinfo{person}{Linyi Yang}, \bibinfo{person}{Tin Lok~James Ng}, \bibinfo{person}{Barry Smyth}, {and} \bibinfo{person}{Riuhai Dong}.} \bibinfo{year}{2020}\natexlab{}.
\newblock \showarticletitle{Html: Hierarchical transformer-based multi-task learning for volatility prediction}. In \bibinfo{booktitle}{\emph{Proceedings of The Web Conference 2020}}. \bibinfo{pages}{441--451}.
\newblock


\bibitem[Yu et~al\mbox{.}(2023)]%
        {yu2023finmem}
\bibfield{author}{\bibinfo{person}{Yangyang Yu}, \bibinfo{person}{Haohang Li}, \bibinfo{person}{Zhi Chen}, \bibinfo{person}{Yuechen Jiang}, \bibinfo{person}{Yang Li}, \bibinfo{person}{Denghui Zhang}, \bibinfo{person}{Rong Liu}, \bibinfo{person}{Jordan~W Suchow}, {and} \bibinfo{person}{Khaldoun Khashanah}.} \bibinfo{year}{2023}\natexlab{}.
\newblock \showarticletitle{FinMem: A performance-enhanced LLM trading agent with layered memory and character design}.
\newblock \bibinfo{journal}{\emph{arXiv preprint arXiv:2311.13743}} (\bibinfo{year}{2023}).
\newblock


\bibitem[Zmandar et~al\mbox{.}(2021)]%
        {zmandar2021joint}
\bibfield{author}{\bibinfo{person}{Nadhem Zmandar}, \bibinfo{person}{Abhishek Singh}, \bibinfo{person}{Mahmoud El-Haj}, {and} \bibinfo{person}{Paul Rayson}.} \bibinfo{year}{2021}\natexlab{}.
\newblock \showarticletitle{Joint abstractive and extractive method for long financial document summarization}. In \bibinfo{booktitle}{\emph{Proceedings of the 3rd Financial Narrative Processing Workshop}}. \bibinfo{pages}{99--105}.
\newblock


\end{thebibliography}

\appendix
\section{Appendix}
\subsection{Prompt Design}
\label{ssec:prompt}
\subsubsection{Prompt for Summarizing Earnings Conference Call Segments}

\begin{itemize}
    \item Identify Key Points \\
    For each segment, identify the key topics covered. Note any significant financial figures, strategic decisions, performance metrics, or forward-looking statements.
    \item Summarize Succinctly \\
    Write a concise summary for each segment, capturing the essence of the discussion. Aim to condense the information into a few sentences that clearly convey the main points and outcomes discussed.
    \item Highlight Relevant Details \\
    Include any specific details that are critical for understanding the segment’s context or implications, such as notable quotes from the company’s executives or specific data points that illustrate trends or changes.
    \item Connect the Dots \\
    If applicable, relate the segment’s content to broader company objectives or industry trends to provide context and show how the segment fits into the bigger picture.
\end{itemize}

\subsubsection{Prompt for Creating an Overview Summary from Earnings Conference Call Segments}
\begin{itemize}
    \item Gather Segment Summaries\\
    Start by reviewing the summaries of each segment from the earnings conference call. Ensure that you have all the segment summaries available to reference.
    \item Identify Common Themes \\
    Look for common themes, recurring issues, or consistent messages across the segments. Note any overarching strategies, goals, or concerns expressed by the company executives.
\end{itemize}

\subsubsection{Prompt for Extracting Important Sentences.}
\begin{itemize}
    \item Check to the Call \\
    Begin by thoroughly listening to the entire earnings conference call. Pay attention to both the prepared remarks and the question-and-answer session.
    \item Identify Focus Points \\
    Identify statements or discussions that involve significant financial metrics, strategic initiatives, new products or markets, regulatory impacts, or any notable shifts in operations. These are potential focus points that could influence investor perceptions and stock prices.
    \item Document Evidence \\
    For each identified focus point, document the exact wording used, the context in which it was discussed, and who discussed it (e.g., CEO, CFO). This will be crucial for accurate interpretation and analysis.
    \item Analyze Impact on Stock Movement \\
    Pre and Post-Analysis: Examine stock price movements immediately before and after the call to capture initial reactions.
    \item Longer-term Impact \\
    Review stock performance in the days or weeks following the call to assess sustained impacts.
    \item Compare with Market Trends \\
    Ensure to factor in overall market conditions and sector movements to isolate the impact of the earnings call from broader market trends.
\end{itemize}

\subsection{Design of Question Bank}
\label{ssec:question}

\begin{table*}[b]
\caption{The designed Question Bank}
\label{tab:company_analysis}
\footnotesize
\onecolumn
\begin{longtable}{|p{1cm}|p{1.28cm}|m{14.2cm}|}
\hline
\textbf{Focus Category} & \textbf{Focus Item} & \textbf{Questions} \\
\hline
\endfirsthead
\hline
\textbf{Focus Category} & \textbf{Focus Item} & \textbf{Questions} \\
\hline
\endhead
\multirow{4}{*}[-1em]{\parbox{1.1cm}{Financial Indicator}} & Dividend & 
\threerowcell{Q1: Did this company pay the investors dividend? \\
Q2: Have there been any increases or decreases in the stock dividends? If yes, what is the rate at which the dividends have been increasing? \\
Q3: What type of dividend did the company pay?} \\
\cline{2-3}
& Revenue & 
\threerowcell{Q1: What was the company's reported revenue for the past quarter, and how does it compare to the same quarter in the previous year? \\
Q2: What factors influenced the company's revenue performance this quarter? \\
Q3: What are the company's revenue forecasts for the upcoming quarters, and what strategies are in place to achieve these targets?} \\
\cline{2-3}
& Return & 
\threerowcell{Q1: What was the company's net profit margin for this quarter, and how has it changed from the previous quarter or year? \\
Q2: What is the company's Return on Equity (ROE) and Return on Assets (ROA) for this period? \\
Q3: How does the company plan to enhance shareholder value in the upcoming periods? Are there any dividends or buybacks planned?} \\
\cline{2-3}
& Earnings & 
\threerowcell{Q1: Have there been any increases or decreases in the earnings? If yes, what is the rate at which the earnings have been increasing or decreasing? \\
Q2: What is the outlook provided by the executives of this company in relation to the future earnings growth? \\
Q3: Are the earnings above or below compared to the expectations? Are they attractive compared to the peers?} \\
\hline
\multirow{3}{*}[-1em]{\parbox{1.1cm}{Employee Manager}} & Salary & 
\threerowcell{Q1: What percentage of the company's total expenses is currently allocated to employee salaries, and how has this changed in response to recent business developments? \\
Q2: How does your company's compensation structure compare with industry standards, particularly in terms of salary, benefits, and bonuses? \\
Q3: What were the average salary increases or decreases across the company this year compared to last year?} \\
\cline{2-3}
& Pension & 
\threerowcell{Q1: What is the current status of the company's pension fund, and what were the major changes to its funding status over the past year? \\
Q2: How does the company manage its pension liabilities, and what strategies are in place to address any underfunded positions? \\
Q3: What are the expected impacts of current pension commitments on the company's future financial performance?} \\
\cline{2-3}
& Management Change & 
\threerowcell{Q1: Have there been any recent changes in the company's key management positions, including the CEO, etc. \\
Q2: What were the reasons behind any recent management changes, particularly in the CEO position? \\
Q3: What impact are the recent management changes expected to have on the company's strategy in the near term?} \\
\hline
\multirow{3}{*}[-1em]{Cost} & Operating Costs & 
\threerowcell{Q1: What were the total operating costs this quarter compared to the previous quarter? \\
Q2: Which factors contributed to any significant changes in operating costs? \\
Q3: How are you managing operating costs in light of current economic conditions?} \\
\cline{2-3}
& Cost of Goods Sold & 
\threerowcell{Q1: How did the Cost of Goods Sold change this quarter, and what were the driving factors behind these changes? \\
Q2: What percentage of revenue does the Cost of Goods Sold represent, and how does this compare to industry norms? \\
Q3: Are there any initiatives in place to reduce Cost of Goods Sold without compromising quality?} \\
\cline{2-3}
& Marketing and Sales Costs & 
\threerowcell{Q1: How much did the company spend on marketing and sales this quarter? \\
Q2: What specific marketing or sales strategies contributed to these costs? \\
Q3: Are there plans to adjust these strategies in the upcoming quarters based on performance?} \\
\hline
\multirow{3}{*}[-1em]{Expansion} & Geographic Expansion & 
\threerowcell{Q1: What specific regions or markets is the company expanding into, and what factors influenced these selections? \\
Q2: What are the initial costs associated with the geographic expansion and strategies are in place to support it? \\
Q3: What is the projected timeline for the new regional operations to reach profitability?} \\
\cline{2-3}
& Product Line Expansion & 
\threerowcell{Q1: What new products is the company introducing, and what consumer or market needs do they aim to address? \\
Q2: How will the introduction of these products impact the company's production costs and overall financial performance? \\
Q3: Are there any expected synergies between the new products and existing products or services?} \\
\cline{2-3}
& Market Segmentation Expansion & 
\threerowcell{Q1: Which new customer segments is the company targeting, and what research supports this strategic direction? \\
Q2: What marketing strategies will be employed to reach these new segments, and what are the anticipated costs? \\
Q3: What are the growth expectations for these new market segments over the next few years?} \\
\hline
Business & Business & 
\threerowcell{Q1: What are the key projects or initiatives currently being undertaken by the company, and provide a brief overview of what each project entails? \\
Q2: How has the performance of these projects compared to the previous quarter, and how do they stand in relation to competitors in the same sector? What has been the market's response to these initiatives? \\
Q3: What are the generated revenues from these projects for the current reporting period, and what potential risks could impact their future performance?} \\
\hline
Future Outlook & Future Outlook & 
\threerowcell{Q1: What are the company's primary strategic goals for the upcoming year, and what key initiatives are planned to achieve these objectives? \\
Q2: How do these future plans align with current industry trends and market demands? \\
Q3: What are the expected financial impacts of these plans on the company's performance in the short and long term?} \\
\hline
\end{longtable}
\end{table*}

\end{document}